# DESIGN AND COMMISSIONING OF A HIGH-LEVEL CONTROL SYSTEM FOR A MEDICAL ISOCHRONOUS CYCLOTRON

P. Hofverberg[†], J.M. Bergerot, J.M. Bruneau, R. Trimaud, Centre Antoine Lacassagne, Nice, France

*Abstract*

MEDICYC (MEDical CYClotron) is an isochronous cyclotron dedicated to radiotherapy which was built and commissioned in Nice, France, in 1990 by a local team aided by experts from CERN. The cyclotron accelerates negative H to a maximum energy of 65 MeV and uses stripping to extract a proton beam. Its primary purpose is treating ocular melanoma by protontherapy but a significant research activity is also present on beam-lines dedicated for this purpose. An extensive refurbishment program of the cyclotron has been started to cope with the end-of-life and/or the obsolescence of several sub-systems. In this context, a new high-level cyclotron control system has been developed and commissioned in 2021-2024. The primary responsibility of the system is the high-level coordination of the source, the RF system, the beam-line and cyclotron magnets, to produce and deliver a beam with a given set of characteristics. A secondary responsibility is the collection, visualization and analysis of sub-system and beam data for monitoring and pre-emptive fault detection. In this contribution, the control system software architecture is presented and the infrastructure on which the systems are deployed is laid out.

## INTRODUCTION

As is often the case with ageing software frameworks, the previous MEDICYC control system had evolved to a large monolithic application which was difficult to comprehend, maintain and critically, to evolve. For a mature system like MEDICYC, this was for a long time not a significant problem since most minor problems had been worked out and sub-systems did no longer evolve. With the decision to start a refurbishment program of the cyclotron, this control system however became a significant concern. The software architecture had to be made more robust and flexible to cope with new requirements and evolutions of sub-systems. A decision to develop a new high-level control system from a clean sheet was therefore taken. The most important requirements on the new system, relative to the those of the preceding system (not repeated here), were:

1. Allow changes to sub-systems and hardware components with as little impact as possible to the high-level control system.
2. Allow the cyclotron to be operated in degraded modes, where some hardware components are temporarily unavailable, reconfigured or interchanged to cope with a problem.
3. Provide extensive data-logging and data-analysis tools for system monitoring and preventive actions.
4. Support the use of several beam-lines and beam characteristics.

† petter.hofverberg@gmail.com

These requirements primarily focus on creating a system that can handle rapid change, weather it is due to hardware evolutions or failures of components. This reflects the reality of an ageing system that is undergoing major updates.

A second large software framework that has been developed in the context of the MEDICYC refurbishment program is the treatment delivery and dosimetry system. This work is however not presented here but is available elsewhere for the interested reader [1].

## SOFTWARE ARCHITECTURE

The single most important software architectural decision that was taken in response to the given requirements was the implementation of the system according to a hierarchical microservice design pattern. This enforces a modular design with clear boundaries, easy to maintain and evolve. The microservices of the system can be grouped in an IOC (Input-Output-Controller) layer, a coordination layer and an GUI layer, shown in Fig. 1, although the layer structure is not strictly enforced and services can communicate with each other across multiple layer boundaries. Each service is run as a separate process and has the responsibility for a single task, with various degree of complexity. Each layer of microservices is responsible for increasingly complex tasks. The bottom layer (IOC layer) implements the control of a single hardware component, such as a beam-line magnet or a beam detector, whereas a microservice in the coordination layer uses an ensemble of IOC services in turn for a single high-level goal. An example could be the task to coordinate the beam-line magnets such that the beam is transported to a given point with a given set of characteristics. The top layer consists of all GUI applications needed for the operation of the cyclotron and has been implemented using the Qt framework [2]. All GUI applications enforce a strict model/view separation and are implemented similarly to any other microservice. The result is that any number of GUIs can be launched independently without interfering with each other.

Microservice domain logic is implemented using hierarchical state machines (Qt's *QStateMachine*). State machines improves significantly the organisation and readability of the software, and greatly aids in ensuring correct behaviour to real-time input.

Communication between microservices is asynchronous and is implemented over DBUS [3] in a publish/subscribe (observer) pattern [4] over the signal/slot system of the Qt framework. Requests or signals are thus passed over a dedicated DBUS daemon running on a communication server. Messages are put in the receiver's event queue (Qt's

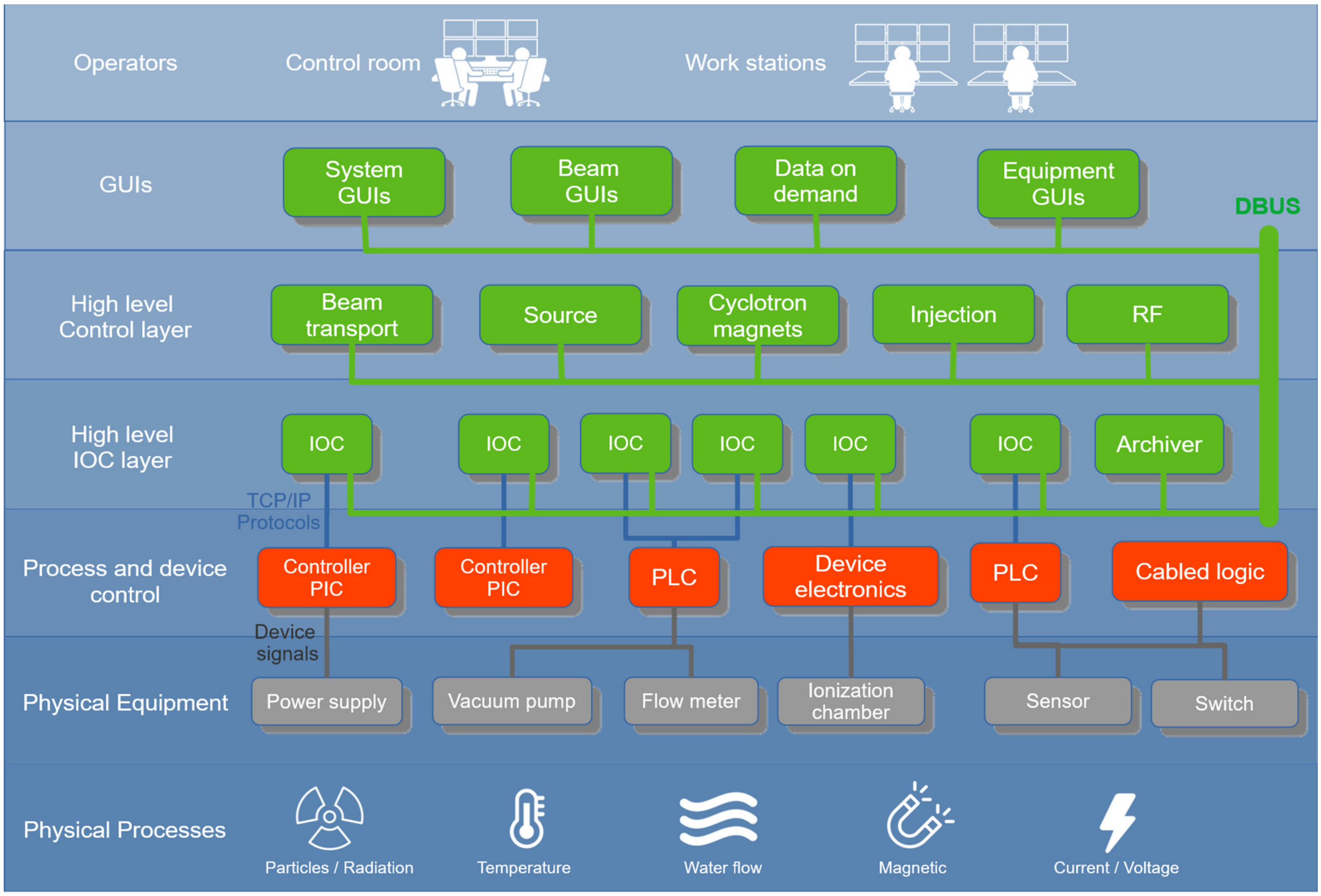

Figure 1: The architecture of the MEDICYC control system, with an emphasis on the high-level components (shown in green color).

*queued connections*) and are treated sequentially in a thread safe implementation. Communication between IOC microservices and low-level hardware (PLCs, micro-controllers, etc) is currently implemented in Omron FINS (Factory Interface Network Service) and various custom written TCP/IP protocols, but an evolution towards OPC-UA (Open Platform Communications Unified Architecture) is however foreseen when a new generation of PLCs (Programmable Logic Controller) are commissioned.

Data logging is done by a dedicated archiver microservice which listens on the DBUS bus for all messages from microservices and sends this to a TimescaleDB database. Applications that require data have then the choice of listening directly over the DBUS for (soft) real-time data, or to query the database.

## DEPLOYMENT ARCHITECTURE

MEDICYC has a longstanding history of using open-source components whenever possible. All computing and network hardware are thus running different flavours of Linux; Debian/Ubuntu for servers, workstations and

Raspberrys, pfSense (FreeBSD) [5] for firewalls, TrueNAS [6] and PBS (Proxmox Backup Server) [7] for backups. In the same spirit, PVE (Proxmox Virtual Environment) [7] was therefore chosen to setup the heart of the control system – the high availability cluster which runs all services necessary to operate the cyclotron (the exception being the graphical user interfaces which runs locally on workstations). The cluster is actually composed of 3 physical nodes (HP Proliant) which communicate with each other over a private network on dedicated NICs. Nodes are multi-master and their configuration is replicated in quasi real-time over the intra-cluster network using incremental OpenZFS send-receive snapshots [8]. The loss of any of the physical nodes thus not result in any service failures. Services are running in light-weight LXC containers and are distributed over the nodes according to the deployment diagram in Fig. 2. All containers are configured as high available services and are restarted and/or migrated automatically on failure, a process that takes less than 60 s. The containers are backed up to the PBS server daily such that it's possible to restore all or part of a container to a given snapshot in the past at any time.

The most important containers and their respective responsibilities are:

1. A communication server, managing all cyclotron microservices and the dedicated Dbus daemon through which the services communicate with each other and with the graphical interfaces. This server thus constitutes the brain of the high-level control of the cyclotron.

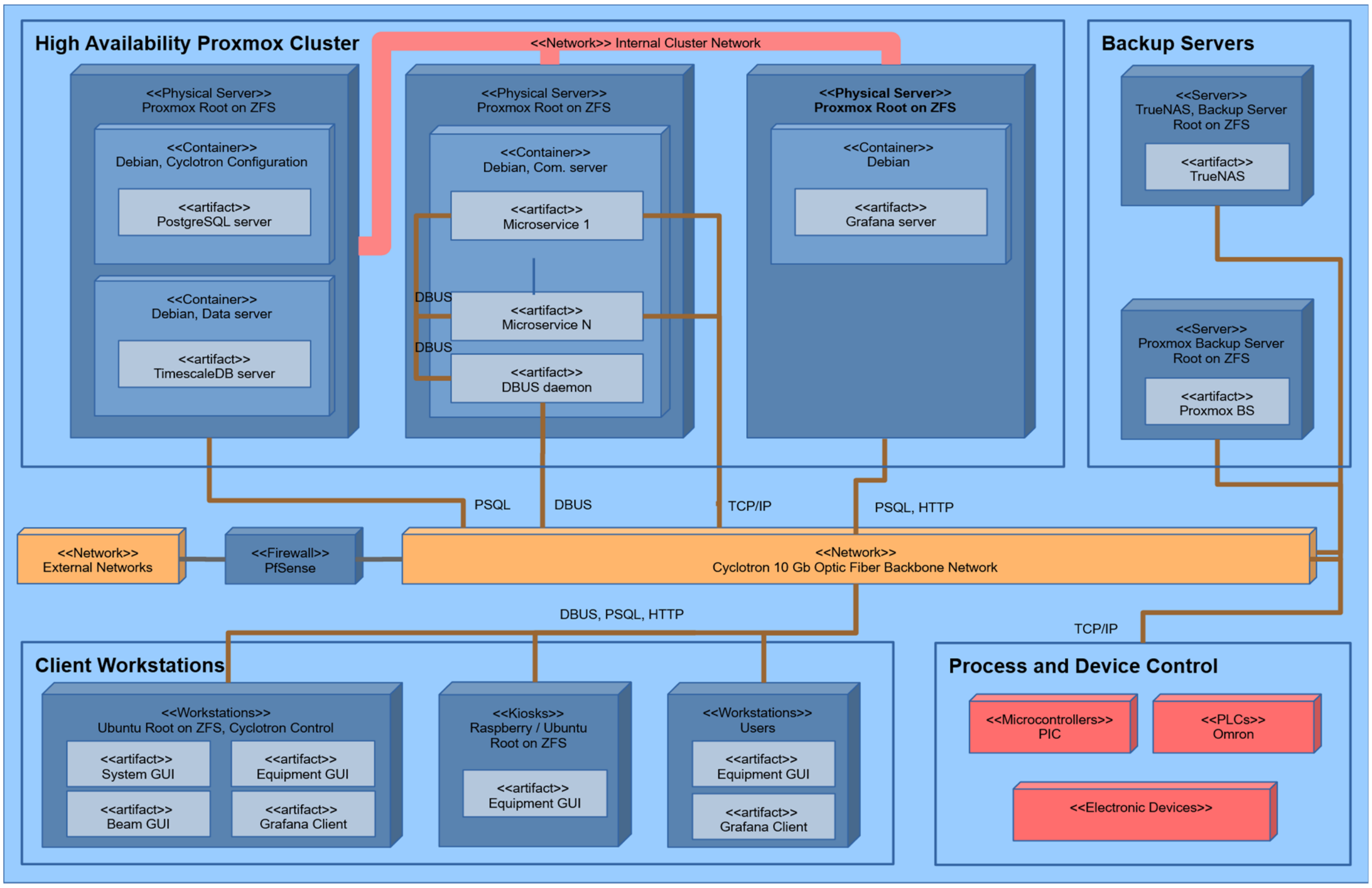

Figure 2: The deployment diagram of the MEDICYC control system.

2. A configuration database server which runs the PostgreSQL database containing the configuration data necessary to run all microservices as well as the set-points and tolerances of all cyclotron sub-systems for all available cyclotron configurations.
3. A timeseries database server, running the TimescaleDB database which stores the timeseries data produced by all microservices.
4. A server running a Grafana web-server through which the timeseries data can be accessed, visualized and interpreted.

A series of Ubuntu DELL workstations are deployed on the internal cyclotron network to host the desktop graphical interfaces required for the operators to run the cyclotron, primarily in the main control room but also on certain workstations throughout the facilities. Raspberrys in touchscreen kiosk mode are also used to control certain hardware. Timeseries data from the Grafana server can be accessed by either these workstations or from user workstations on other networks to monitor the cyclotron at all times and for predictive and posteriori fault detection. The deployed computing hardware operates on a 10 Gb fiber optic backbone network in a loop topology for high availability.

## CONCLUSIONS

A new distributed high-level control system has been developed and commissioned for the MEDICYC cyclotron. An emphasis on modularity and flexibility was made to allow for the ongoing upgrades of the cyclotron and their sub-system to proceeded smoothly. The implemented hierarchical microservice software architecture has proven to respond well to these requirements. Focus is now shifting to the underlying control system layers, where PLCs and micro-controllers are being replaced by modern versions.